\documentclass[twocolumn,showpacs,preprintnumbers,amsmath,amssymb,aps,prb,superscriptaddress]{revtex4-2}
\usepackage{epsfig}
\usepackage{graphicx}% Include figure files
\usepackage{dcolumn}% Align table columns on decimal point
\usepackage{bm}% bold math
\usepackage{color} % \color{red} \color{black}

\begin{document}

\title{
Bulk-Sensitive Spin-Resolved Resonant Electron Energy-Loss Spectroscopy (SR-rEELS): \\
Observation of Element- and Spin-Selective Bulk Plasmons
}
\author{Shin-ichi Kimura}
\email{kimura@fbs.osaka-u.ac.jp}
\affiliation{Graduate School of Frontier Biosciences, Osaka University, Suita, Osaka 565-0871, Japan}
\affiliation{Department of Physics, Graduate School of Science, Osaka University, Toyonaka, Osaka 560-0043, Japan}
\affiliation{Institute for Molecular Science, Okazaki, Aichi 444-8585, Japan}
\author{Taishi Kawabata}
\affiliation{Department of Physics, Graduate School of Science, Osaka University, Toyonaka, Osaka 560-0043, Japan}
\author{Hiroki Matsumoto}
\affiliation{Department of Physics, Graduate School of Science, Osaka University, Toyonaka, Osaka 560-0043, Japan}
\author{Yu Ohta}
\affiliation{Department of Physics, Graduate School of Science, Osaka University, Toyonaka, Osaka 560-0043, Japan}
\author{Ayuki Yoshizumi}
\affiliation{Department of Physics, Graduate School of Science, Osaka University, Toyonaka, Osaka 560-0043, Japan}
\author{Yuto Yoshida}
\affiliation{Department of Physics, Graduate School of Science, Osaka University, Toyonaka, Osaka 560-0043, Japan}
\author{Takumi Yamashita}
\affiliation{Department of Physics, Graduate School of Science, Osaka University, Toyonaka, Osaka 560-0043, Japan}
\author{Hiroshi Watanabe}
\affiliation{Graduate School of Frontier Biosciences, Osaka University, Suita, Osaka 565-0871, Japan}
\affiliation{Department of Physics, Graduate School of Science, Osaka University, Toyonaka, Osaka 560-0043, Japan}
\author{Yoshiyuki Ohtsubo}
\affiliation{Graduate School of Frontier Biosciences, Osaka University, Suita, Osaka 565-0871, Japan}
\affiliation{Department of Physics, Graduate School of Science, Osaka University, Toyonaka, Osaka 560-0043, Japan}
\author{Naoto Yamamoto}
\affiliation{High Energy Accelerator Research Organization (KEK), Tsukuba, Ibaraki 305-0801, Japan}
\author{Xiuguang Jin}
\affiliation{High Energy Accelerator Research Organization (KEK), Tsukuba, Ibaraki 305-0801, Japan}
\date{\today}
\begin{abstract}
We have developed a spin-resolved resonant electron energy-loss spectroscopy (SR-rEELS) in the primary energy of 0.3--1.5~keV,
which corresponds to the core excitations of $2p$-$3d$ absorption of transition metals and $3d$-$4f$ absorption of rare-earths, with the energy resolution of about 100~meV using a spin-polarized electron source as a GaAs/GaAsP strained superlattice photocathode.
Element- and spin-selective carrier and valence plasmons can be observed using the resonance enhancement of core absorptions and electron spin polarization.
Furthermore, bulk-sensitive EELS spectra can be obtained because the primary energy corresponds to the mean free path of 1--10~nm.
The methodology is expected to provide us novel information of elementary excitations by resonant inelastic x-ray scattering and resonant photoelectron spectroscopy.
\end{abstract}

%
%\pacs{}% PACS, the Physics and Astronomy
%%%%%%%%%%%%%%%%%%%%%%%%%%%%%%%%%%%%%%%%%%%%%%%%%%%%%%%%%%%%
\maketitle
%
%%%%%%%%%%%%%%%%%%%%%%%%%%%%%%%%%%%%%%%%%%%%%%%%%%%%%%%%%%%%
\section{Introduction}
In recent years, resonant PhotoElectron Spectroscopy (rPES) and Resonant Inelastic X-ray Scattering (RIXS) using synchrotron radiation
are widely used for element-specific observation methods of elementary excitations of interacting electrons, so-called quasiparticles,
and collective excitations such as magnons and plasmons~\cite{Ament2011}.
These methods have been successful in observing physical quantities, such as the direct observation of the hybridization bands between conduction and localized $4f$ states in rare-earth compounds~\cite{Im2008} and magnon dispersions in high-$T_c$ cuprate superconductors~\cite{Ishii2021}.

%%%%%%%%%%%%%% FIG. 1. Resonant Spectroscopies %%%%%%%%%%%%%%%%%%%%
\begin{figure*}[t]
\begin{center}
\includegraphics[width=0.8\textwidth]{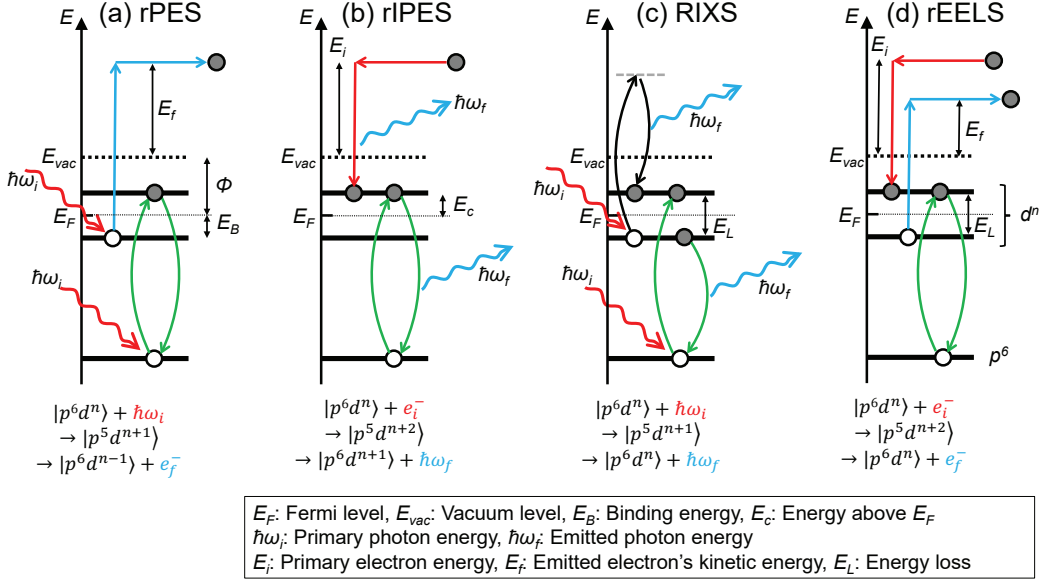}
\end{center}
\caption{
Resonant high-energy spectroscopies using synchrotron radiation and monochromatized electron source, 
(a) resonant PhotoElectron Spectroscopy (rPES), 
(b) resonant Inverse PhotoElectron Spectroscopy (rIPES),
(c) Resonant Inelastic X-ray Scattering (RIXS),
and (d) resonant inelastic electron scattering, namely resonant Electron Energy-Loss Spectroscopy (rEELS).
The electronic configuration is an example of the resonance of the $2p$-$3d$ absorption edge of transition metal compounds.
In the case of rare-earth compounds, the absorption process should be changed to the $3d$-$4f$ absorption.
These methods provide complementary information of quasiparticles and collective excitations of solids.
}
\label{fig:Spectroscopies}
\end{figure*}
%%%%%%%%%%%%%%%%%%%%%%%%%%%%%%%%%%%%%%%

As examples of transition metal compounds, the rPES and RIXS optical processes depicted in Figs.~\ref{fig:Spectroscopies}(a) and \ref{fig:Spectroscopies}(c) can be explained as follows:
\begin{eqnarray}
{\rm rPES:} |p^6 d^n\rangle + \hbar\omega_i &\to& |p^5 d^{n+1}\rangle \to |p^6 d^{n-1}\rangle + e^-_f  \nonumber \\
{\rm RIXS:} |p^6 d^n\rangle + \hbar\omega_i &\to& |p^5 d^{n+1}\rangle \to |p^6 d^{n}\rangle + \hbar\omega_f  \nonumber
\end{eqnarray}
Here, $|p^6 d^n\rangle$ is the ground state, $n$ the number of $d$ electrons, $|p^5 d^{n+1}\rangle$ the intermediate state after the $2p$-$3d$ absorption, and $\hbar\omega_i$, $\hbar\omega_f$, and $e^-_f$ the incident photon, emitted photon, and emitted electron, respectively.
Complementary to these techniques are resonant Inverse PhotoElectron Spectroscopy [rIPES, Fig.~\ref{fig:Spectroscopies}(b)], 
which is the inverse process of rPES, 
and resonant Electron Energy-Loss Spectroscopy [rEELS, Fig.~\ref{fig:Spectroscopies}(d)], which is the inelastic electron scattering. 
Electronic structures are resonantly excited by a monochromatic electron beam at energies of inner-shell absorptions in both methods.
The optical processes of rIPES and rEELS are as follows:
\begin{eqnarray}
{\rm rIPES:} |p^6 d^n\rangle + e^-_i \to |p^5 d^{n+2}\rangle &\to& |p^6 d^{n+1}\rangle + \hbar\omega_f  \nonumber \\
{\rm rEELS:} |p^6 d^n\rangle + e^-_i \to |p^5 d^{n+2}\rangle &\to& |p^6 d^{n}\rangle + e^-_f  \nonumber
\end{eqnarray}
Here, $|p^5 d^{n+2}\rangle$ is the intermediate state after the resonant excitation by the primary electron $e^-_i$ at the energy of the $2p$-$3d$ absorption.
Although rPES and rIPES are opposite processes, they can detect the occupied and unoccupied final electronic structure where the number of electrons increases by one or decreases by one from that of the ground state of $|p^6 d^n\rangle$, respectively.
On the other hand, the final state of rEELS is the same as RIXS, 
but the intermediate state is different.
Therefore, the comparison of rEELS spectra to RIXS spectra will provide a deep insight into the intermediate states.

In comparison with RIXS, rEELS has some advantages as follows:
The first is to cover a wide momentum (wavenumber) range owing to the shorter de Broglie wavelength of electrons than that of photons~\cite{Ibach2009a},
the second to measure the spin-polarization of electrons directly,
and the third to achieve atomic-scale spatial resolution using an electron microscope technique.
However, rEELS at the $2p$-$3d$ absorption edges of transition metal compounds and the $3d$-$4f$ absorption edges of rare-earth compounds 
($0.3 - 1.5$~keV) has not been performed yet. 
In contrast, many studies have been done using RIXS at these absorption edges.
Also, some rEELS experiments have been performed only with the primary energy ($E_i$) of 
less than about 150~eV, which corresponds to the $3p$-$3d$ absorption edges of transition metal compounds and the $4d$-$4f$ absorption edges of rare-earth compounds~\cite{Modesti1985,DellaValle1989,Gorschluter1994,Fromme1995a,Gorschluter1998}.
However, the energy resolution of rEELS is currently about a few hundred of meV, which is poorer than the current standard resolutions of rPES 
($\sim 10$~meV in the vacuum-ultraviolet region~\cite{Kimura2010} and $\sim 100$~meV in the soft-x-ray region~\cite{Kimura2014a}) 
and RIXS ($\sim 100$~meV)~\cite{Strocov2010}.
One of the reasons is that it is difficult to change the energy while maintaining good energy resolution and high electron flux with a commonly used combination of thermal electron sources and monochromators.

There are two types of EELS techniques widely used at present.
One is High-Resolution EELS (HR-EELS) using low-energy electron beams of $E_i \leq 100~{\rm eV}$ 
with the energy resolution of less than several tens of meV, which is widely used as a method for observing 
the $E-\vec{q}$ curve ($\vec{q}$ is the momentum transfer) of collective excitations such as plasmons and phonons of solid surfaces~\cite{Vig2017}.
The other is TEM-EELS, which uses high-energy electron beams of several tens of keV or higher using a transmission electron microscope~\cite{Hofer2016}.
TEM-EELS is used to measure the transmitted EELS of ultra-thin samples
and has the advantage of obtaining bulk and thin-film's specific information in a small area by using an electron microscope.
However, except for a few advanced experiments~\cite{Krivanek2019}, 
the typical energy resolution is not high (several hundreds of meV)~\cite{Grogger2008}.
Also, since the electron beam energy is fixed, it is difficult to perform element-specific measurements using resonance.
It is also difficult to observe spin-selective spectra.

So far, Spin-Resolved EELS (SR-EELS) has been performed to observe spin waves and magnons of transition metal thin films~\cite{Plihal1999,Tang2007,Prokop2009,Vollmer2003,Zakeri2012,Chuang2012,Zakeri2013a}.
The energy resolution is as high as that of HR-EELS because the HR-EELS technique is used~\cite{Ibach2003}.
Also, Spin-Resolved resonant EELS (SR-rEELS) has been performed to observe $d$-$d$ and $f$-$f$ excitations of transition metal oxides and rare-earth oxides~\cite{Fromme1994,Fromme1995,Fromme1995a,Fromme2001}.
In this case, the energy resolution is about a few hundred eV due to the changing the primary energy.
These previous works have been performed using primary electrons with $E_i \leq 100~{\rm eV}$, which electrons are very surface sensitive 
owing to the shorter mean free path than one nm~\cite{Zangwill1988}.
To have bulk sensitivity, the mean free path of electrons must be longer than one nm,
so the primary electron energy must be higher than several hundreds of eV~\cite{Zangwill1988}.

In this paper, a newly-developed SR-rEELS instrument with a spin-polarized highly-monochromatized electron source 
of $E_i = 0.3-1.5$~keV is reported.
The purpose is to observe element-specific, spin-resolved elementary excitations in the bulk of solids, 
mainly topological materials such as Dirac and Weyl semimetals and strongly correlated electron systems such as high-$T_c$ materials and heavy fermions.
This method is complementary to HR-EELS and TEM-EELS and will provide additional information to rPES and RIXS.
First, the setup is explained, followed by the characteristics of the electron source and its energy resolution.
Then, some EELS spectra for bulk-sensitivity, resonance effect, and spin-selectivity are reported.

%%%%%%%%%%%%%%%%%%%%%%%%%%%%%%%%%%%%%%%%%%%%%%%%%%
\section{Spin-resolved resonant electron-energy-loss spectroscopy (SR-rEELS) apparatus}
%
%%%%%%%%%%%%%% FIG. 2. rEELS apparatus %%%%%%%%%%%%%%%%%%%%
\begin{figure*}[t]
\begin{center}
\includegraphics[width=0.7\textwidth]{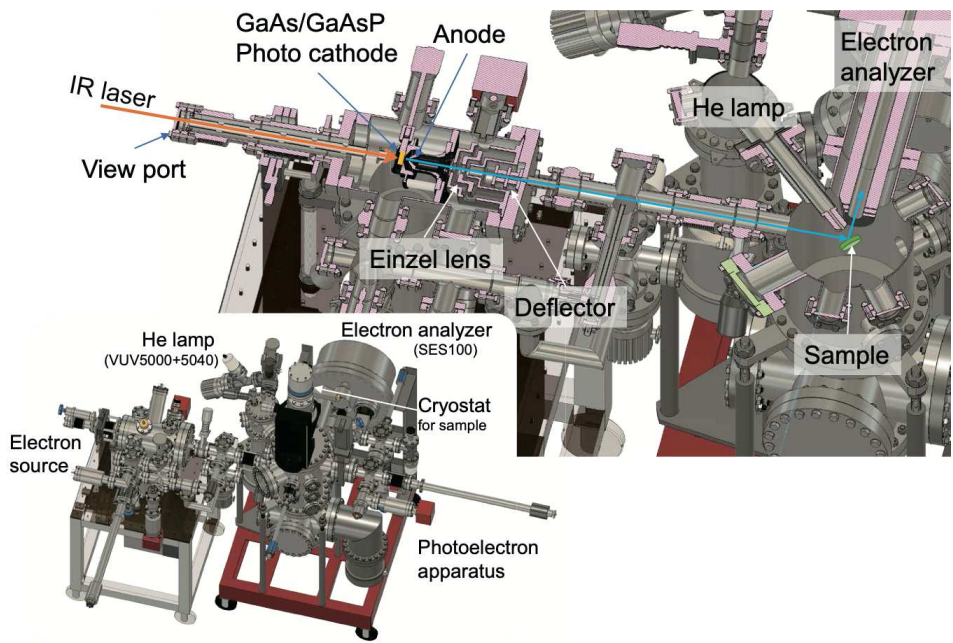}
\end{center}
\caption{
Spin-Resolved resonant Electron Energy-Loss Spectroscopy (SR-rEELS) setup.
The overview of the SR-rEELS apparatus, including spin-polarized electron source and photoelectron spectroscopy setup, 
is shown in the lower figure.
The electron path of the incident and scattered electrons 
with the IR laser of the excitation to the photocathode in the cross-sectional view of the setup is also shown in the upper figure.
The photoelectron setup equips a hemispherical electron analyzer (SES100) 
and two light sources, a He lamp with a monochromator (VUV5000+5040) and an x-ray tube (XR-50, not shown in the figure).
}
\label{fig:apparatus}
\end{figure*}
%%%%%%%%%%%%%%%%%%%%%%%%%%%%%%%%%%%%%%%
Figure~\ref{fig:apparatus} shows an overview and a cross-sectional view of the SR-rEELS apparatus and the electron beam path.
The SR-rEELS apparatus is a combination of a photoelectron spectrometer and a spin-polarized electron source.

The spin-polarized electron source is a negative electron affinity (NEA) surface of GaAs/GaAsP strained superlattices 
with the spin polarization of more than 80~\%~\cite{Saka2000,Yamamoto2008}, 
which is much higher than that of ordinary bulk GaAs ($\sim50$~\%)~\cite{Pierce1980a}.
Two kinds of GaAs/GaAsP strained superlattices were used~\cite{Jin2008}:
One is a 12-pair strain-compensated superlattice (SCSL), and the other a 12-pair strained superlattice (SSL)~\cite{Jin2013a}.
The NEA surface is created by depositing Cs on the GaAs/GaAsP surface.
An IR laser of 785~nm wavelength (L785P090, THORLABS Co.) is irradiated from the backside through the GaP substrate to the GaAs surface to emit electrons~\cite{Yamamoto2008}.
The generated electron beam is focused on the sample surface by an Einzel lens and a deflector that adjusts the XY axis.
This electron source is used for SR-rEELS and Spin-Resolved Reflection High-Energy Electron Diffraction (SR-RHEED) measurements of the same sample condition,
so the photocathode is designed to be capable of applying up to $-30$~kV, 
and the constant voltage supply with an extreme-high voltage accuracy of less than one ppm (IPES$-30$kV, MBScientific AB) is used for acceleration.
The acceleration energies (bias voltages to the photocathode) for SR-rEELS and SR-RHEED are 0.3--1.5~keV ($-0.3--1.5$~kV) and 20--30~keV ($-20--30$~kV), respectively.

To obtain a spin-polarized electron beam,
the incident linearly-polarized IR laser light is converted to left and right circular polarization using a 1/4-$\lambda$ plate and irradiated to the photocathode.
The sample is then irradiated with spin-polarized electron beams
whose electron spins are parallel/antiparallel to the direction of the incident electrons.
The electron spin will be manipulated by a Wien filter~\cite{Yasue2014} in the future.

In order to irradiate a sample with a highly monochromatized electron beam and easily control the acceleration energy, 
electrons near the threshold energy of the NEA surface of GaAs/GaAsP were used.
To evaluate the energy width of the electron beam, we introduced the electron beam to the electron analyzer directly
and evaluated the energy width by observing the spectral distribution.

The photoelectron spectrometer consists of a hemispherical analyzer (SES100, VG-Scienta AB) with an orbital radius of 100~mm and
two light sources:
One is a helium discharge lamp with a monochromator (lamp: VUV5000, monochromator: VUV5040, VG-Scienta AB) and the other an x-ray tube (XR-50, SPECS GmbH).
The purpose of the photoelectron spectrometer except for EELS is to compare the quasiparticle electronic structure with obtained EELS spectra
and check the surface conditions
by angle-resolved photoelectron spectroscopy and x-ray photoelectron spectroscopy.

%%%%%%%%%%%%%% FIG. 3. Resolution %%%%%%%%%%%%%%%%%%%%
\begin{figure}[t]
\begin{center}
\includegraphics[width=0.40\textwidth]{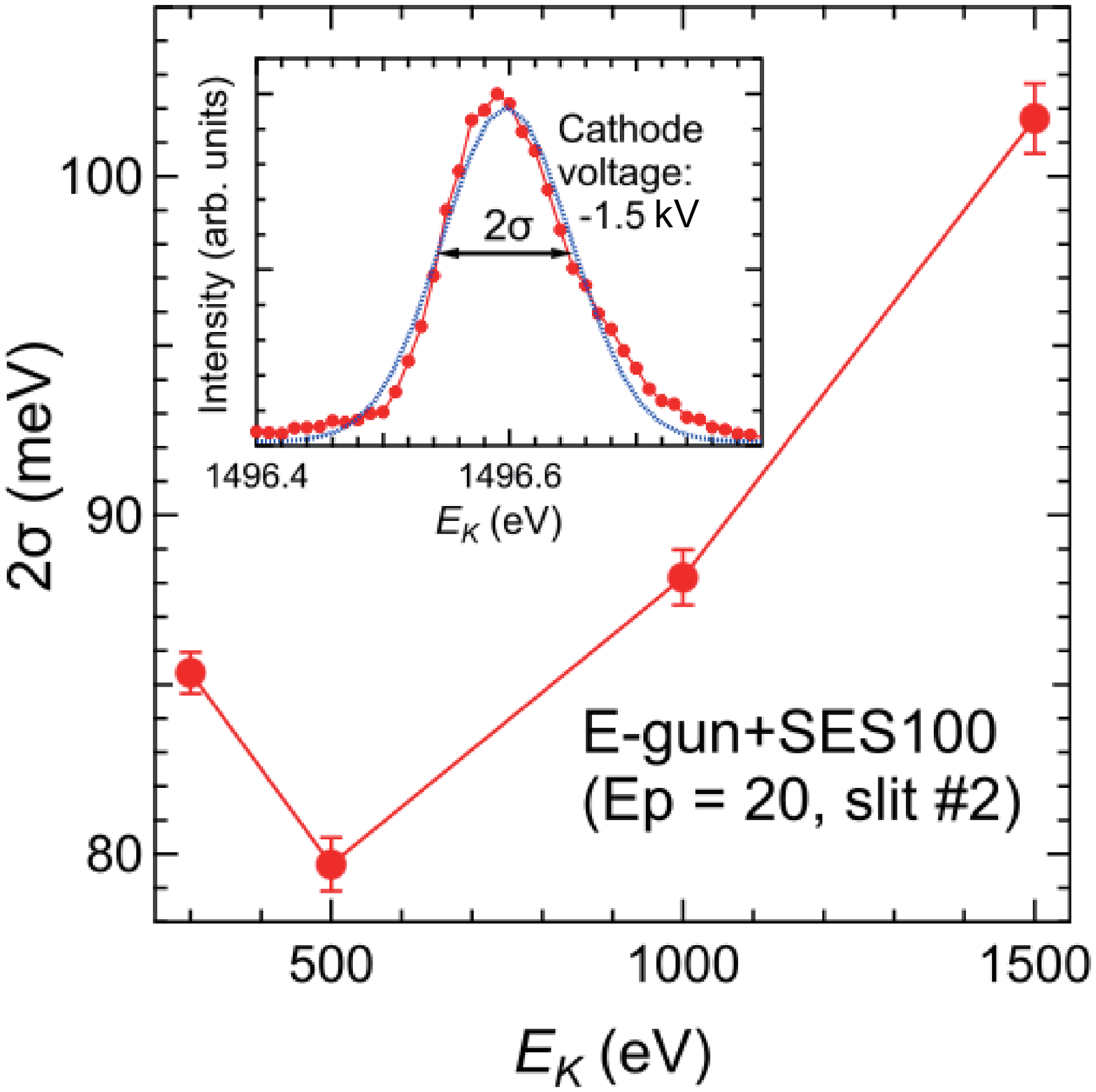}
\end{center}
\caption{
Energy width ($2\sigma$) of the electron source as a function of kinetic energy.
The electron analyzer's energy resolution is set as about 20~meV by the pass energy $E_p$ and slit number.
Inset: A typical spectrum of the electron beam of $E_K=1500$~eV from the electron source directly measured by the photoelectron analyzer.
}
\label{fig:resolution}
\end{figure}
%%%%%%%%%%%%%%%%%%%%%%%%%%%%%%%%%%%%%%%
The inset of Fig.~\ref{fig:resolution} is a typical spectrum showing the energy distribution of the electron beam 
with the kinetic energy of 1.5~keV (the bias voltage of the photocathode was set as $-1.5$~kV) from the electron source.
SSL has been used as the photocathode.
The observed peak width was $101.7~\pm~1.0$~meV for $2\sigma$ evaluated by fitting with Gaussian.
The maximum beam current and electron flux was roughly evaluated as about sub $\mu$A and $10^{12}$~s$^{-1}$ 
measured by a gold-evaporated aluminum plate at the laser power of about 100~mA.
The beam size at the sample position is about 200~$\mu$m in diameter,
so the obtained brightness is about $3\times10^{13}~{\rm s^{-1}mm^{-2}}$.
The kinetic energy dependence of the peak width is shown in the main figure of Fig.~\ref{fig:resolution}.
From this figure, $2\sigma$ is 80--85~meV at $E_K=300$ and 500~eV, and as $E_K$ increases, $2\sigma$ increases.
Here, the energy resolution of the electron analyzer was set as about 20~meV.
So, the energy width of electron beam is about 77~meV (100~meV) at $E_K=500$~eV (1500~keV).
The minimum energy width observed at $E_K=500$~eV is similar to that of the photoluminescence (PL) peak~\cite{Jin2013a}.
Since the PL peak of SCSL is sharper than that of SSL, the energy width of SCSL is expected to become narrower than that shown in Fig.~\ref{fig:resolution}.

Although the energy width obtained here is not as narrow as that of HR-EELS~\cite{Ibach2017}
or very high energy-resolution TEM-EELS ($\sim10$~meV)~\cite{Krivanek2014},
the energy width is much narrower than that previously reported, 
about 0.25~eV and more of typical EELS at $E_i\sim1$~keV~\cite{Gorschluter1994, Lounis2008} 
and of SP-rEELS at $E_i\sim100$~eV~\cite{Fromme1995a,Gorschluter1998}.
The reason is considered to use the GaAs/GaAsP photocathode.

The obtained energy width is approximately equivalent to the room temperature energy width of $4k_{\rm B}T \sim 0.1$~eV.
In other words, thermal fluctuations of electrons in the photocathode are considered to be manifested.
It has been reported that the energy width becomes narrower when the photocathode is cooled down~\cite{Pastuszka2000}.
Therefore, to further narrow the energy width, the photocathode needs to be cooled.

The present energy resolution of the SR-rEELS instrument is almost consistent with 
the present standard soft-x-ray-RIXS (SX-RIXS) beamlines~\cite{Strocov2010,Harada2012},
so SR-rEELS becomes a complementary method to the SX-RIXS.
Furthermore, since SR-rEELS can characterize the spin-polarization of electrons,
it can provide additional information of electron spins to the present SX-RIXS.

%%%%%%%%%%%%%%%%%%%%%%%%%%%%%%%%%%%%%%%%%%%%%%%%%%
%%%%%%%%%%%%%%%%%%%%%%%%%%%%%%%%%%%%%%%%%%%%%%%%%%
\section{Measurement results}
\subsection{EELS on gold and silver films}
%
%%%%%%%%%%%%%% FIG. 4. rEELS apparatus %%%%%%%%%%%%%%%%%%%%
\begin{figure}[t]
\begin{center}
\includegraphics[width=0.48\textwidth]{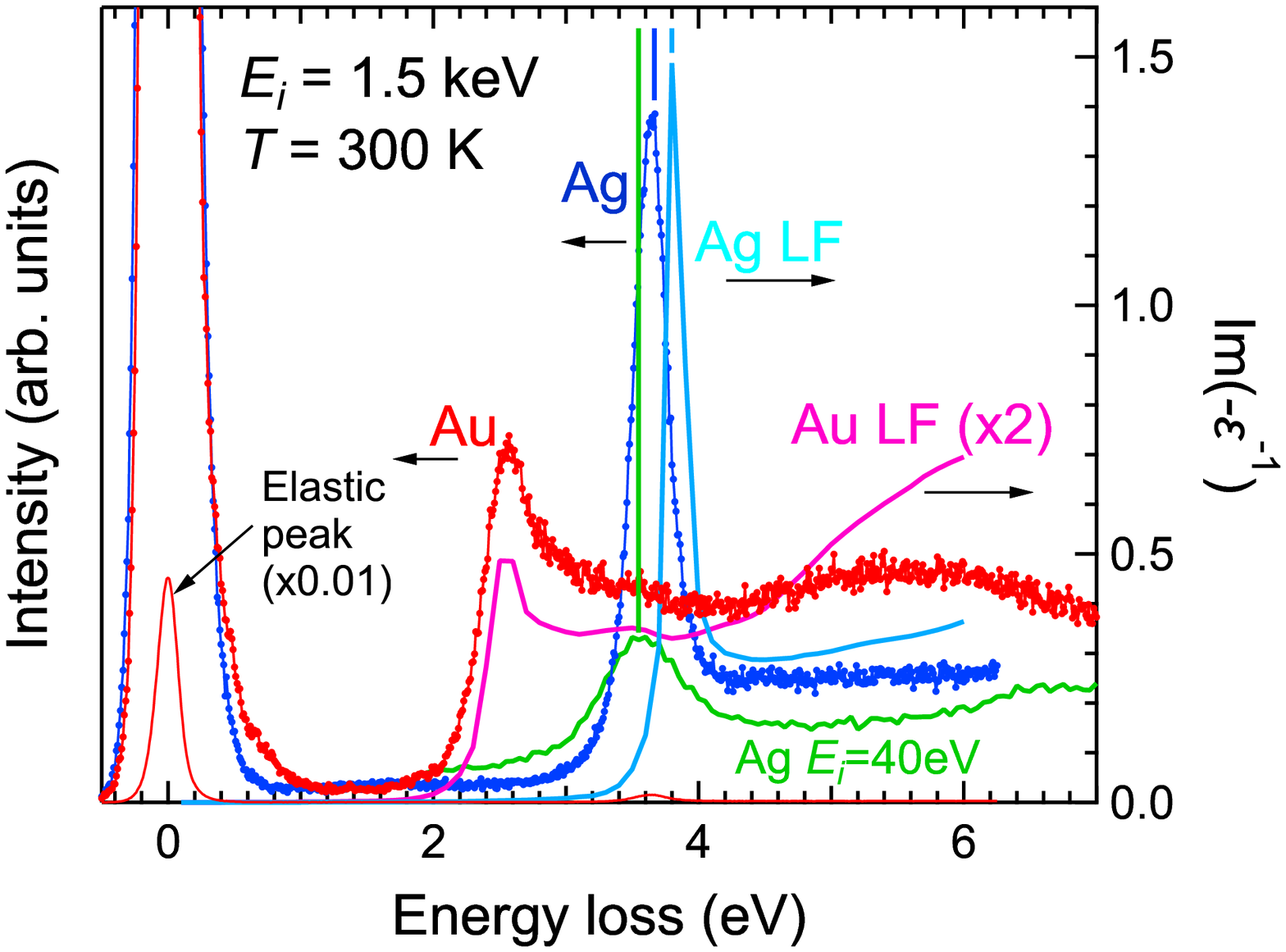}
\end{center}
\caption{
EELS spectra of gold (Au) and silver (Ag) films at $E_i = 1.5$~keV at the temperature of 300~K accompanied with corresponding loss function (LF) spectra obtained by the Kramers-Kronig analysis of optical reflectivity spectra~\cite{Babar2015}, and low-energy EELS (LE-EELS) spectra of an Ag film ($E_i=40$~eV).
The LF peaks of Au (Energy loss: $E_L\sim2.5$~eV) and Ag ($E_L\sim3.8$~eV) indicate the bulk plasmons.
The LE-EELS peak ($E_L\sim3.55$~eV) of Ag references the surface plasmon.
Vertical lines are the energies of plasmon peaks of Ag measured by the three methods.
}
\label{fig:AuAg}
\end{figure}
%%%%%%%%%%%%%%%%%%%%%%%%%%%%%%%%%%%%%%%
In order to evaluate the SR-rEELS instrument,
EELS spectra of gold (Au) and silver (Ag) polycrystalline thin films at $E_i = 1.5$~keV were measured as shown in Fig.~\ref{fig:AuAg}.
The loss function spectra [$Im(-\varepsilon(\omega))^{-1}$: LF] derived from the Kramers-Kronig analysis of the optical reflectivity spectra~\cite{Babar2015} 
as a reference for a bulk sensitive experiment
and the low-energy EELS (LE-EELS) spectrum of Ag with $E_i = 40$~eV as a reference for a surface-sensitive experiment are shown in the figure.
In the EELS spectrum of Au, carrier plasmon peaks at the energy loss $E_L\sim2.5$~eV, 
and valence plasmon peaks at $E_L\sim3.5$~eV and 5.5~eV are visible.
The peaks at $E_L\sim2.5$~eV and 3.5~eV are in good agreement with the LF spectrum.
The 5.5-eV peak does not appear in the LF spectrum owing to the limitation of the energy range.
Since the LF spectrum reflects the bulk information obtained in the optical reflectivity spectrum,
i.e., it can be regarded as a bulk plasmon spectrum,
the peaks of the EELS spectrum are also attributed to the bulk plasmons.
This result can be regarded as a benefit of using the bulk-sensitive 1.5-keV electron beam.

The EELS spectrum of the Ag film with $E_i = 1.5$~keV, on the other hand, shows a carrier plasmon peak at $E_L\sim3.65$~eV.
In comparison with the bulk plasmon peak at $E_L\sim3.8$~eV in the LF spectrum 
and the surface plasmon peak at $E_L\sim3.55$~eV in the LE-EELS spectrum, 
the peak energy of 1.5-keV EELS spectrum is located at intermediate energy.
The low-energy side's tail of the 1.5-keV EELS peak at $E_L \sim 3$~eV is almost consistent with the LE-EELS peak, 
and the high-energy side of the peak (the kink of the spectrum) at $E_L \sim 4$~eV is consistent with that of the LF peak.
This result suggests that the 1.5-keV EELS spectrum includes both the surface and bulk components.

\subsection{rEELS on a transition metal compound NiO}
%
%%%%%%%%%%%%%% FIG. 5. NiO rEELS %%%%%%%%%%%%%%%%%%%%
\begin{figure}[t]
\begin{center}
\includegraphics[width=0.40\textwidth]{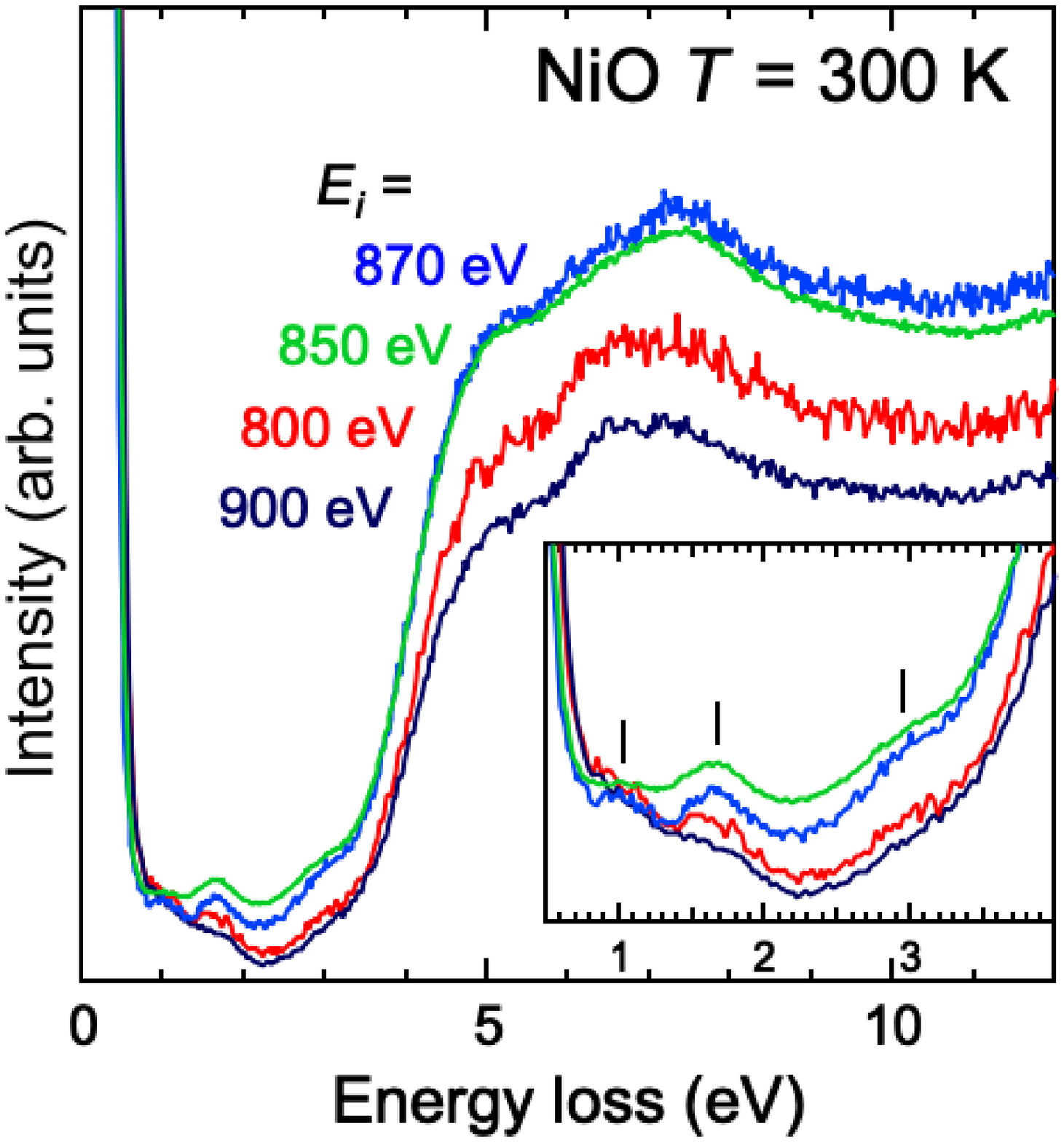}
\end{center}
\caption{
Resonant EELS spectra of nickel monoxide NiO with the primary energies ($E_i$) of 800--900~eV.
The overall spectral intensity is enhanced at $E_i$ = 850 and 870~eV near the Ni~$2d$-$3d$ absorption edge.
(Inset) Enlarged view of the Ni~$d$-$d$ excitations (indicated by vertical lines) below the lowest charge transfer excitation energy of about 4~eV.
}
\label{fig:NiO}
\end{figure}
%%%%%%%%%%%%%%%%%%%%%%%%%%%%%%%%%%%%%%%
The primary energy dependence of the EELS spectra of nickel monoxide (NiO) at the Ni~$2p$-$3d$ absorption edge is reported 
to check the resonance effect in the EELS measurement. 
Several previous SR-EELS studies have been performed at $E_i \leq 100$~eV~\cite{Fromme1995a, Jones2000, Gorschluter1994, Fromme1994}.
The rEELS spectra at $E_i=800-900$~eV near the Ni $2p$-$3d$ absorption edge are shown in Fig.~\ref{fig:NiO}.
These spectra were normalized at the intensity of the elastic peak.
The main structure in the energy region above 4~eV originates from the charge transfer excitation 
from the hybridization band between the Ni~$3d$ lower-Hubbard band and the O~$2p$ state 
to the Ni~$3d$ upper-Hubbard band~\cite{Powell1970,Sawatzky1984,Michels1986}.
The intensity of the spectrum at $E_i=900$~eV is slightly lower than that at 800~eV due to decreasing the photoionization cross-section of the Ni~$3d$ state~\cite{Yeh1985}.
Compared to the spectra at $E_i=800$~eV and 900~eV, the overall spectral intensities at $E_i=850$~eV and 870~eV increase.
Since the Ni $2p_{5/2}$ and $2p_{3/2}$ absorption edges are located at about 850~eV and 870~eV, 
the spectral intensity enhancements are considered to originate from the resonance effect of the Ni~$2p$-$3d$ absorption.
It should be noted that the enhancement ratio is not as high as that of the rPES at the $2p$ absorption edge~\cite{Tjernberg1996}.
The reason for the low resonance effect in rEELS is not clear yet, but this is the first observation of rEELS at the $2p$-$3d$ absorption edge.

The inset of Fig.~\ref{fig:NiO} shows the Ni~$d$-$d$ excitation peaks below the charge-transfer gap at about 4~eV.
Among these peaks, especially the peaks at $E_L\sim1.7$~eV and 3.0~eV are resonantly enhanced.
This result is also in good agreement with the rEELS result~\cite{Gorschluter1998} for the Ni~$3s$-edge ($\sim$~100~eV).
In the optical reflectivity spectrum of the sample, there is no peak in this energy range~\cite{Powell1970}
because the $d$-$d$ excitations are not dipole-allowed transitions.
Therefore, the rEELS can provide additional information for the forbidden transition to optical spectra.

%%%%%%%%%%%%%%%%%%%%%%%%%%%%%%%%%%%%%%%
%%%%%%%%%%%%%%%%%%%%%%%%%%%%%%%%%%%%%%%
\subsection{rEELS on a rare-earth compound SmB$_6$}
%
%%%%%%%%%%%%%% FIG. 6. SmB6 rEELS %%%%%%%%%%%%%%%%%%%%
\begin{figure}[t]
\begin{center}
\includegraphics[width=0.42\textwidth]{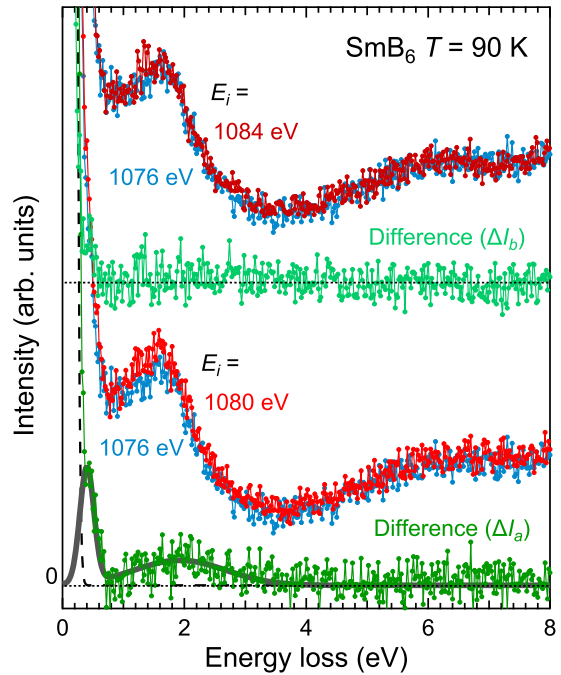}
\end{center}
\caption{
EELS spectra of SmB$_6$ with the primary energies $E_i$ of 1080, 1084, and 1087~eV and difference spectra 
[$\Delta I_a=I(E_i=1080~{\rm eV})-I(E_i=1076~{\rm eV})$ and $\Delta I_b=I(E_i=1084~{\rm eV})-I(E_i=1076~{\rm eV})$].
The dashed line is a Gaussian fitting curve to the rest elastic peak of $\Delta I_a$.
The solid line is a fitting curve with two Gaussians after subtracting the dashed line from $\Delta I_a$.
The lower-energy peak at about 0.4~eV and the higher-energy peak at about 2~eV are considered to originate from 
the $4f$-$5d$ interband transition and the carrier plasmon, respectively.
In the $\Delta I_b$ spectrum, the 0.4-eV peak is slightly visible, but the 2-eV peak disappears.
}
\label{fig:SmB6}
\end{figure}
%%%%%%%%%%%%%%%%%%%%%%%%%%%%%%%%%%%%%%%
Next, we introduce the resonance enhancement of carrier plasmons of samarium hexaboride (SmB$_6$), 
which has recently attracted attention as a topological Kondo insulator~\cite{Li2020b,Dzero2010,Dzero2016,Miyazaki2012,Ohtsubo2019}.
SmB$_6$ is a typical valence fluctuation material in which Sm$^{3+}$ and Sm$^{2+}$ ions coexist.
This carrier originates from the Sm$^{3+}$~$5d$ band and is expected to have a resonance enhancement at the Sm$^{3+}$~$3d$-$4f$ absorption edge ($E_i\sim1080$~eV)~\cite{Kaindl1984}.
The peak energy of the Sm$^{2+}$~$3d$-$4f$ absorption is about 1076~eV, which is slightly lower than that of the Sm$^{3+}$ ion.
Therefore, the EELS spectrum at $E_i$=1080~eV is expected to be different from those at $E_i$=1076~eV and other energies.

Figure~\ref{fig:SmB6} shows the EELS spectra at $E_i=1076$~eV (Sm$^{2+}$ peak), 1080~eV (Sm$^{3+}$ peak), 
and 1084~eV (the higher-energy tail of the Sm$^{3+}$ $3d$-$4f$ absorption), 
and the difference spectra of $\Delta I_a=I(E_i=1080~{\rm eV})-I(E_i=1076~{\rm eV})$ 
and $\Delta I_b=I(E_i=1084~{\rm eV})-I(E_i=1076~{\rm eV})$.
Peaks at $E_L\sim2$ and 6~eV are observed in all EELS spectra.
Looking at the $\Delta I_a$ spectrum, the peak at $E_L\sim2$~eV is resonantly enhanced at $E_i=1080$~eV.
The 2-eV peak originates from carrier plasmons 
because the plasma edge of the reflectivity spectrum of this material is located at the same energy~\cite{Kimura1994}.
On the other hand, the peak at $E_L\sim6$~eV is not enhanced.
This result implies that the 6-eV peak originates from the electronic excitation in the boron $2s$ and $2p$ orbitals, 
not directly related to the Sm$^{3+}$ site~\cite{Kimura1992a}.
Therefore, the carrier plasmon is resonantly enhanced at the Sm$^{3+}$~$3d$-$4f$ absorption edge, i.e., it is confirmed to originate from the Sm$^{3+}$ site.
It should be noted that the 2-eV peak is not visible in the $\Delta I_b$ spectrum.
This result suggests that the enhancement of the 2-eV peak is due to the Sm$^{3+}$ $3d$-$4f$ absorption, rather than a monotonic change in the photoionization cross-section.

After subtracting the elastic peak by a Gaussian fitting from the $\Delta I_a$ spectrum, 
the 2-eV carrier plasmon peak and a peak at 0.4~eV become visible as indicated by a solid thick line (fitted by using two Gaussians).
In the $\Delta I_b$ spectrum, the subtraction of the spectrum at the peak energy of the Sm$^{2+}$ $3d$-$4f$ absorption from that at the higher energy of the Sm$^{3+}$ absorption, even though the 2-eV peak does not appear, the 0.4-eV peak is slightly visible.
This result suggests that the 0.4-eV peak has a dominant Sm$^{3+}$ component than the 2-eV peak.
In the energy region at around 0.4~eV, the Sm~$4f$-$5d$ absorption has been observed in the optical conductivity spectrum~\cite{Kimura1994}.
Therefore, this is reasonable evidence that
the resonant enhancement of the spatially expanded carrier plasmon is smaller than that of the interband transition between localized electronic states.
Combining with the result of NiO, it is suggested that the resonant enhancement is considered to be strongly related to the localization of electrons.
The relation of the resonant enhancement to the electron localization should be investigated systematically.

%%%%%%%%%%%%%%%%%%%%%%%%%%%%%%%%%%%%%%%
%%%%%%%%%%%%%%%%%%%%%%%%%%%%%%%%%%%%%%%
\subsection{SR-EELS on a magnetized iron film}
%
%%%%%%%%%%%%%% FIG. 7. Fe SR-EELS %%%%%%%%%%%%%%%%%%%%
\begin{figure}[t]
\begin{center}
\includegraphics[width=0.42\textwidth]{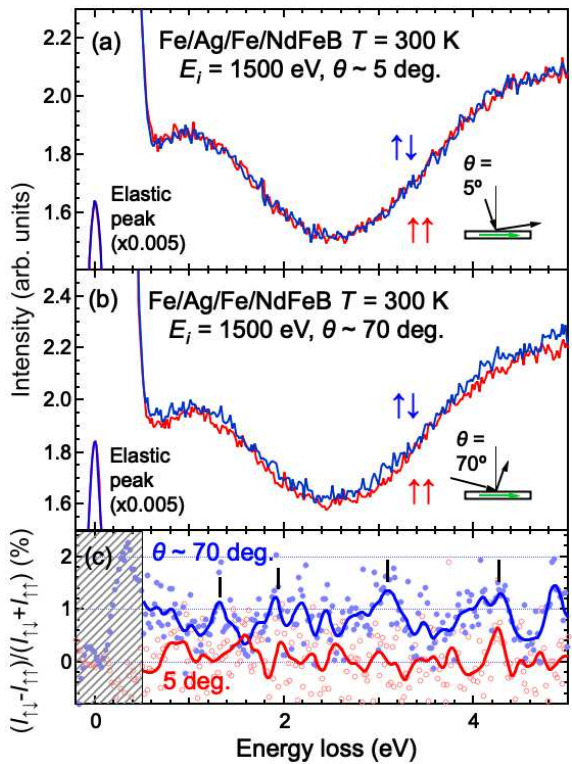}
\end{center}
\caption{
Spin-resolved EELS spectra of a magnetized iron (Fe) film taken at the incident angles of 5~deg. (a) and 70~deg. (b) from the sample normal direction,
and their spin asymmetry [$(I_{\uparrow \downarrow}-I_{\uparrow \uparrow})/(I_{\uparrow \downarrow}+I_{\uparrow \uparrow})$] spectra (c).
The spin asymmetry spectra (solid circles for 70~deg., open circles for 5~deg.) are very noisy due to the very low inelastic scattering intensities, 
so the spectra are smoothed by averaging every ten points as shown by solid lines.
Solid vertical lines indicate observed peaks.
The hatched area is a strongly effective region of the elastic peak, so the data are unreliable.
}
\label{fig:Fe}
\end{figure}
%%%%%%%%%%%%%%%%%%%%%%%%%%%%%%%%%%%%%%%
Finally, to check the spin-polarized inelastic scattering, 
the SR-EELS using a magnetized iron (Fe) thin film has been measured shown in Fig.~\ref{fig:Fe}.
Thanks to the itinerant ferromagnetic property of Fe at room temperature, 
the spin-polarized band structure is induced by a weak external magnetic field.
Then, spin-selective inelastic scatterings due to the spin-polarized electronic structure are expected.

To observe SR-EELS on Fe at the primary energy region as high as one keV, we prepared an induced-ferromagnetic Fe film:
A Fe substrate with a size of about $2 \times 2 \times 0.1$~mm$^3$ attached to the top of an Nd-Fe-B magnet ($\sim 1 \times 1 \times 1$~mm$^3$).
The direction of magnetization was set parallel to the substrate surface in the incident plane.
The magnetization intensity at the substrate surface was about 10$^{-4}$~T.
The silver polycrystalline thin film with a thickness of a few tens of nm was deposited on this substrate as a buffer layer,
and a few nm of Fe polycrystalline film was deposited on the top of it in a high vacuum less than $1\times10^{-6}$~Pa.
Here, the Fe film is not a single crystal.
However, this is not a problem because the purpose here is to observe the change in SR-EELS spectra
due to the difference of the scattering intensity of spin-polarized electrons by the magnetization direction.

The spin-polarized electrons were obtained by injecting a left-and-right circularly polarized IR laser beam into a photocathode
with a 1/4-$\lambda$ plate tilted by $\pm45$~deg. from the linear polarization plane.
The spin directions of primary electrons are parallel or antiparallel to the incident electron beam direction.

Figure~\ref{fig:Fe} shows the spectra with the incident electron spin 
parallel ($\uparrow \uparrow$) and antiparallel ($\uparrow \downarrow$) to the magnetization direction. 
The angle between the incident electron beam and the surface perpendicular is set to 
a near-normal incidence geometry of 5~deg. (Fig.~\ref{fig:Fe}(a)) and a grazing incidence geometry of 70~deg. (Fig.~\ref{fig:Fe}(b)) 
from the normal direction of the sample surface 
with the constant scattering deviation of 90~deg. (see Fig.~\ref{fig:apparatus})
The spectra are normalized at the elastic peak intensity.
In the near-normal incidence geometry of Fig.~\ref{fig:Fe}(a), almost identical spectra were obtained for the two combinations.
This result is because the directions of the incident electron spin and the magnetization are almost orthogonal, 
and there is little change even when the electron spin direction is changed.
The result also implies that the sample magnetization does not affect to the electron beam directly.

On the other hand, in Fig.~\ref{fig:Fe}(b), where the incident electron spin direction is almost parallel to the magnetization direction,
the intensity of $\uparrow \downarrow$ is slightly higher than that of $\uparrow \uparrow$ in the whole $E_L$ region above 0.5~eV.
The spin asymmetry [$(I_{\uparrow \downarrow}-I_{\uparrow \uparrow})/(I_{\uparrow \downarrow}+I_{\uparrow \uparrow})$] spectra 
of different incident angles are shown in Fig.~\ref{fig:Fe}(c).
The spectrum of the near-normal incidence (5~deg.) is almost zero, which suggests no magnetic anisotropy again.
In contrast, the spectrum of the grazing incidence (70~deg.) has an asymmetry of about 1~\% in the whole region.
After smoothing the asymmetry spectra by averaging every ten points, 
some peaks at $E_L\sim$1.3, 1.9, 3.1, and 4.3~eV are visible (indicated by vertical solid lines).
This result is almost consistent with that of a previous SR-EELS spectrum for 20~ML bcc-Fe/Ag(100)~\cite{Komesu2006} and Fe/W(110)~\cite{Samarin2015}.
Therefore, the SR-EELS at $E_i = 1.5$~keV can provide information on the spin-polarization of electronic structures.

%%%%%%%%%%%%%%%%%%%%%%%%%%%%%%%%%%%%%%%
%%%%%%%%%%%%%%%%%%%%%%%%%%%%%%%%%%%%%%%
\section{Conclusion}

We have developed spin-resolved resonant electron energy-loss spectroscopy (SR-rEELS) at the primary energy region of 0.3--1.5~keV, 
corresponding to the transition metals' $2p$-$3d$ and rare-earths' $3d$-$4f$ absorption edges.
The current energy resolution is about 100~meV or less. 
Resonant enhancements of the charge transfer and $d$-$d$ excitations of NiO, 
and the carrier plasmon and the $4f$-$5d$ absorption of SmB$_6$ have been observed.
Furthermore, spin-dependent EELS spectrum parallel and antiparallel to the direction of the electron beam were observed in a magnetized Fe film.
The SR-rEELS instrument will be helpful for the detection of spin currents, in which the spin direction is fixed in the direction of the current, 
and the physics originating from electron spins.

%%%%%%%%%%%%%%%%%%%%%%%%%%%%%%
\section*{Acknowledgments}
We would like to thank Prof. Iga for providing high-quality SmB$_6$ samples.
This work was partly supported by JSPS KAKENHI (Grant Nos. 15H03676, 20H04453), and The Research Grants of Mitsubishi Foundation, 
The Murata Science Foundation, and The Research Foundation for Opto-Science and Technology.

%%%%%%%%%%%%%%%%%%%%%%%%%%%%%%
\section*{Data Availability}
The data that support the findings of this study are available from the corresponding author upon reasonable request.

%%%%%%%%%%%%%%%%%%%%%%%%%%%%%%
%\begin{thebibliography}{99}
%
%
%\end{thebibliography}

%\bibliographystyle{plain}
\bibliographystyle{apsrev4-1}
\bibliography{../../../bibtex/library}

\end{document}